\documentclass{article}

\usepackage[english]{babel}
\usepackage{amsfonts}
\usepackage{amsmath}
\usepackage{amssymb}
\usepackage{amsthm}
\usepackage{array}
\usepackage{multirow}
\usepackage{float}
\usepackage{booktabs}
\usepackage{color}

\usepackage{authblk}
\usepackage{hyperref}
\usepackage{graphicx}

\newcommand{\CHAIN}[1]{\mathbf{#1}}

\renewcommand{\P}{{\CHAIN{P}}}
\newcommand{\Q}{{\CHAIN{Q}}}
\newcommand{\U}{{\CHAIN{U}}}
\newcommand{\PO}{{\CHAIN{P_0}}}
\newcommand{\QO}{{\CHAIN{Q_0}}}

\renewcommand{\b}{\beta}

\newcommand{\dtau}{d\tau}

\newcommand{\dual}{\overline}

\def\ea{a}
\def\eb{b}
\def\ec{c}
\def\ex{x}

\newcommand{\PF}{\mathbf{\Pi_{free}}}

\newcommand{\p}{d \! p \,}
\newcommand{\q}{d \! q \,}
\newcommand{\dpo}{d \! p_0 \,}
\newcommand{\dqo}{d \! q_0 \,}

\newcommand{\dt}{d \! t \,}
\newcommand{\dx}{d \! x \,}
\newcommand{\dto}{d \! t_0 \,}
\newcommand{\dxo}{d \! x_0 \,}

\newcommand{\ddp}{\delta_p}
\newcommand{\ddq}{\delta_q}
\newcommand{\ddtau}{\delta_\tau}

\newcommand{\PrPO}{Pr(\PO)}
\newcommand{\PrQO}{Pr(\QO)}
\newcommand{\Np}{N_\P}
\newcommand{\Nq}{N_\Q}
\newcommand{\rp}{r_\P}
\renewcommand{\rq}{r_\Q}
\newcommand{\rpo}{r_\PO}
\newcommand{\rqo}{r_\QO}

\newtheorem{definition}{Definition}

\begin{document}
\title{An Introduction to Influence Theory:\\ Kinematics and Dynamics}

\author[1]{\small Kevin H. Knuth}
\author[1]{\small James L. Walsh}

\affil[1]{\footnotesize Department of Physics, University at Albany (SUNY), Albany NY, USA}


\maketitle

\abstract{
  Influence theory is a foundational theory of physics that is not based on traditional empirically defined concepts, such as positions in space and time, mass, energy, or momentum.  Instead, the aim is to derive these concepts, and their empirically determined relationships, from a more primitive model.  It is postulated that there exist things, which we call particles, that influence one another in a discrete and directed fashion resulting in a partially ordered set of influence events.  We consider the problem of consistent quantification of the influence events.  Observers are modeled as particle chains (observer chains) as if an observer were able to track a particle and quantify the influence events that the particle experiences.  From these quantified influence events, we study consistent quantification of the universe of events based on the observer chains. In this paper we both review and further develop the kinematics and dynamics of particles from the perspective of influence theory.
}


\section{Introduction}

This paper presents an introduction to influence theory, which is based on an elementary model of physical interactions, in which objects influence one another in a discrete and directed fashion, from which a foundational theory of physics emerges as the unique quantified description of influence events.  In developing a foundational theory, aimed partly at expanding the explanatory power of theoretical physics, one does not have the liberty to base the theory on empirically defined concepts, such as positions in space and time, mass, energy, or momentum.  Instead, the aim is to derive these concepts, and their empirically determined relationships, from a more primitive, foundational, model.

Here, within the context of kinematics and dynamics, we summarize our previously published work describing how concepts of space and time emerge from the theory as the unique quantified description of events \cite{Knuth+Bahreyni:JMP2014}, noting other previously published work describing how the Dirac equation for the free particle (in 1+1 dimensions) emerges from the theory \cite{Knuth:Info-Based:2014} as the result of probabilistic inferences \cite{Knuth+Skilling:2012, GKS:PRA, Skilling+Knuth:MPQ}, and we present new results focused on the action of a free particle and the emergence of the concepts of force and potential energy in the case of a weakly influenced particle.

\section{Relationality and the Foundation of Mechanics}

Classical mechanics is one of the first great triumphs of physics and, as such, it has a firm place at the foundations of physics.  Its focus is on the physics of macroscopic objects, for which we have developed an intuition, which involves concepts such as position, time, velocity, acceleration, force, mass, energy, and momentum.  Since a solid training in physics instills a degree of familiarity with these concepts, it is often not appreciated how difficult it was to develop this intuition over the last several hundred years \cite{Jammer:2009:Mass, Jammer:2012:Force}, or how we continue to do so \cite{Tolman:1912, Lehrman:1973, Okun:1989, Sonego+Pin:2004, Espinoza:2005, Hecht:2009, Jammer:2009:Mass, Jammer:2012:Force}. At this point in history, one might not expect to encounter foundational work focused on kinematics and dynamics, even in the relativistic sense.  However, a new foundational theory must be demonstrated to consistently explain well-known experimental results.  That is what we aim to accomplish in this work.

%
Despite the great success of mechanics and its reconciliation with our intuition, or perhaps more accurately, the re-education of our intuition, it became clear with the introduction of relativity that mechanics required some significant revision when pushed outside the realm of our everyday experience.  The fact that space and time are related to one another, and the fact that lengths, distances, and durations are observer-dependent cast serious doubts on our perception of reality.  In many ways, both the theory of relativity and the theory of quantum mechanics made it clear that much of physics is observer-centric \cite{Wiener:1936:Observer, Cramer:1988, Goldstein:1998}.
%
%
%
It need not be that reality is observer-dependent; it simply must be that these quantities used to describe macroscopic objects in actuality describe the relationships between those objects and an observer.  Precisely what such relationships represent is unknown---this is the hidden physics.

Much of the physics describing the behavior of an object relies on parameters that we often consider to be properties possessed by that object.  However, many of these ``properties'' are observer-dependent, which means that they cannot represent anything possessed by the object. Instead, they must somehow represent the relationship between the object and the observer.  This is known as \emph{relationality} \cite{Rovelli:1996, Rovelli:2017}.  Physics describes interactions, and the interactions between objects in the physical world and observers, which are also objects in the physical world, make physics a participatory sport \cite{Wheeler:1990, Zeilinger:2004, Fuchs:2017}.  However, we are mindful of Jaynes' \emph{Mind Projection Fallacy} \cite{Jaynes:1990:ProbabilityLogic}, and recognize the fact that our knowledge about the universe, which is gained through our interactions with it, is not the same as the universe itself.

Our goal has been to develop an elementary model of physical interactions in which the familiar concepts of mechanics (more specifically, relativistic mechanics)  emerge as observer-dependent quantities that describe the relationships between objects and observers.  This work presented here, which we call \emph{influence theory}, is an early step toward this goal.

\section{Influence and Influence Events}

\subsection{Mathematical Preliminaries}

We begin with some basic order-theoretic definitions involving partially-ordered sets, or posets.  The interested reader is encouraged to consult a detailed text \cite{Birkhoff:1967, Gratzer2003, Davey&Priestley} for more information.
\begin{definition}
A \textbf{partially ordered set} $(P,\leq)$, or \textbf{poset}, is a set of elements $P$ along with a binary ordering relation, generally denoted $\leq$, which, for elements $\ea, \eb,$ and $\ec \in P$, satisfies
\begin{align}
P1. \quad & \ea \leq \ea & \; \mbox{(Reflexivity)} \nonumber \\
P2. \quad  & \;\mbox{if}\; \ea \leq \eb \;\mbox{and}\; \eb \leq \ea \;\mbox{then}\; \ea = \eb  & \; \mbox{(Antisymmetry)} \nonumber \\
P3. \quad  & \;\mbox{if}\; \ea \leq \eb \;\mbox{and}\; \eb \leq \ec \;\mbox{then}\; \ea \leq \ec & \; \mbox{(Transitivity).} \nonumber
\end{align}
For all elements $\ea, \eb \in P$ we have that either $\eb$ \emph{includes} $\ea$, denoted $\ea \leq \eb$, or $\ea$ includes $\eb$, denoted $\eb \leq \ea$, or $\ea$ and $\eb$ are incomparable, denoted $\ea \; || \; \eb$.  It is for this reason, that there possibly exist pairs of elements that cannot be ordered, that $(P, \leq)$ is called a \emph{partially} ordered set.
\end{definition}
For elements $\ea$ and $\eb \in P$ where $\ea \leq \eb$ and $\ea \neq \eb$, we write $\ea < \eb$.  In addition to \emph{inclusion}, we will also utilize the concept of \emph{covering}.  Given two poset elements $\ea,\eb \in P$ for which $\ea \neq \eb$, we say that $\eb$ \emph{covers} $\ea$, denoted $\ea \prec \eb$, if and only if $\eb$ includes $\ea$, $\ea < \eb$, and there does not exist any element $\ex$ such that $\ea < \ex < \eb$.

A totally-ordered set, or a chain, is a special case of a partially-ordered set.
\begin{definition}
A \textbf{totally-ordered set}, or a \textbf{chain}, is a poset $P$ for which given any two elements $\ea$ and $\eb \in P$ either $\ea \leq \eb$ or $\eb \leq \ea$.  That is, there are no incomparable elements in the chain.
\end{definition}
With these definitions in hand, we proceed to introduce the model.

\subsection{The Influence Model}

We postulate that there exist observer-independent objects that interact with, or at least influence, each other and that people possess means (i.e., equipment) to know something about these interactions.  We begin with a simplified model that strips down these ideas to a minimal few, and we work to discover where this leads us.
It would be naive to expect that this present effort will result in a final theory, but we are hopeful that this first attempt at a foundational theory will demonstrate feasibility while providing new insights on how to proceed.

Inspired by Feynman and Wheeler's absorber theory \cite{Wheeler+Feynman:1945, Wheeler+Feynman:1949}, which was an attempt to do away with the abstract concept of a field, we imagine that there exist objects, which we will call particles for lack of a better name.  We \emph{postulate that particles can influence one another in a directed and discrete fashion}.  We could choose to assume that particles interact with one another in a symmetric and bi-directional fashion.  However, we believe that directed asymmetric influence, for which one particle influences a second particle, is a simpler assumption.  It is important to note that this is largely a matter of preference, with each postulate leading to a different theory.  However, a bidirectional and symmetric interaction is difficult to imagine from a foundational perspective, since two objects would have to somehow coordinate and agree on a choice to interact \cite{Cramer:1988, Kastner:2012}. Instead, we assume that one object influences another and that the mechanism that governs this is essentially unpredictable, i.e. random, with each particle having an equal probability of being influenced in any given instance of influence, possibly subject to constraints not yet identified.  Since one particle influences, and the other particle is influenced, an act of influence is not symmetric.  Despite this asymmetry, this picture of influence still contains some features of what one might consider to be an interaction since in an instance of an act of influence, the first particle influences the second particle and not any other.  This fact may have consequences for the first particle as well as the second, and a resulting physics that exhibits symmetry with respect to the particles' identity remains a possibility.

\begin{figure}
\centering
\makebox{\includegraphics[width=1\columnwidth]{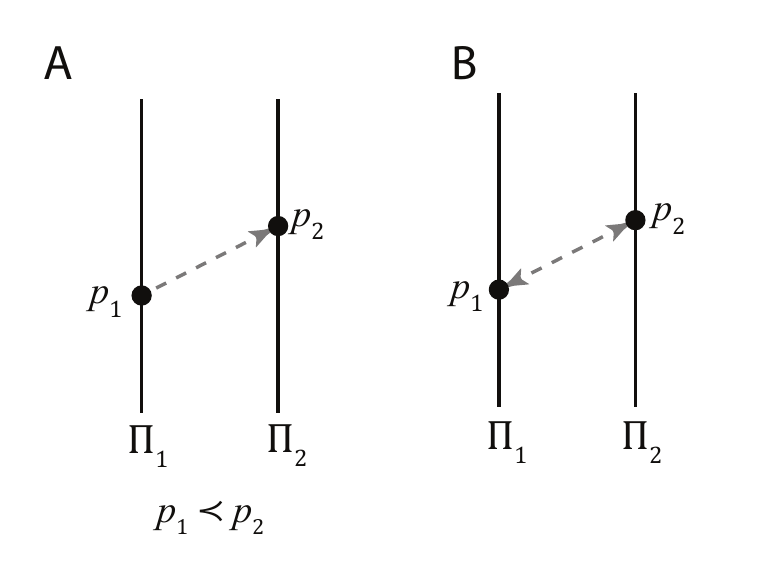}}
\caption{A. This figure illustrates directed influence where particle $\Pi_1$ influences particle $\Pi_2$.  This defines two events $p_1 \in \Pi_1$ and $p_2 \in \Pi_2$, which can be ordered $p_1 \prec p_2$. B. This figure illustrates mutual interaction.  The symmetry of interaction precludes the ability to order the two events $p_1 \in \Pi_1$ and $p_2 \in \Pi_2$. In this case, one does not obtain a partially-ordered set of events.} \label{fig:influence-vs-interaction}
\end{figure}

The second, and perhaps more relevant, reason for modeling influence as a directed act is that by considering both the act of influencing experienced by the first particle and the act of being influenced experienced by the second particle as two disparate events, we can assign to them a partial order.
That is, the act of a particle $\Pi_1$ influencing another particle $\Pi_2$ results in two distinct events $p_1 \in \Pi_1$ and $p_2 \in \Pi_2$, such that $p_1 \prec p_2$,\footnote{It should be noted that this direction of the ordering, $p_1 \prec p_2$, is an arbitrary convention.} where $p_1$ represents the act of $\Pi_1$ influencing $\Pi_2$, and $p_2$ represents the act of $\Pi_2$ being influenced by $\Pi_1$.
A symmetric interaction would mean that
the two related events, $p_1$ and $p_2$, would have an ambiguous order-theoretic relationship to one another jeopardizing the partial order.
The asymmetric directional influence results in a partially ordered set of events, which is basically a form of pre-time, which would be jeopardized by a symmetric influence
(Figure \ref{fig:influence-vs-interaction}).

Last, we \emph{postulate that the events experienced by a given particle can be totally ordered}.  It is likely that there are multiple ways in which such a postulate can be justified.  For example, we have shown that postulating that each influence event experienced by a particle couples two internal states, and postulating that each internal state couples two influence events leads to a total order of alternating internal states and influence events \cite{Knuth:Info-Based:2014}.  Here we will simply postulate that events experienced by each particle are totally ordered, since the internal states are assumed to be inaccessible, and are thus otherwise irrelevant.

The result is that we describe the set of influence events as a partially ordered set (poset), and the particles as totally-ordered chains of events.  However, since influence refers to the interaction between two particles, this theory is only concerned with situations in which there are multiple particles.  In fact, the influence events experienced by, and used to describe, the particle are the result of interactions with other particles.  This is clarified by the concept of a \emph{pchain} (short for particle chain).
%
\begin{definition}
A \textbf{pchain} $\Pi$ is a totally ordered set of events (a chain) such that each event $p_i \in \Pi$ is paired with another event $q_i \in \Pi_{i}$ on a pchain $\Pi_{i}$ via an act of influence, such that either $p_i$ covers $q_i$, $p_i \succ q_i$ or $q_i$ covers $p_i$, $p_i \prec q_i$.   A pchain is a special case of a chain.
\end{definition}
In this paper, it will be assumed that $\Pi$ and $\Pi_{i}$ are distinct particle chains, that is, there is no possibility of self-influence.\footnote{The implications of self-influence will be studied in a future work.}

A pchain will be used to model a particle.  We will often refer to a \emph{particle chain}, where it is understood that the particle chain is a pchain.
It should be noted that this definition in conjunction with the assumption that there is no self-influence implies the fact that the existence of one pchain implies the existence of other pchains.  That is, you cannot have a universe with only one particle.  This is reasonable since we are considering universes in which there are observers to study and describe the universe in which they are embedded.

Each event on a pchain represents an act of influence.  For this reason, we generically refer to events comprising a pchain as \emph{influence events} (despite the fact that there is no other kind of event).  At times, it is of some benefit to distinguish events on a pchain that result from that pchain influencing another pchain from events that result from that pchain being influenced by another pchain.  This leads to the pchain-centric definitions of an \emph{influence event} and an \emph{outfluence event}.
\begin{definition}
An \textbf{influence event} $p$ on a pchain $\Pi$, $p \in \Pi$, is an event that represents the act of being influenced.
\end{definition}
\begin{definition}
An \textbf{outfluence event} $p$ on a pchain $\Pi$, $p \in \Pi$, is an event that represents the act of influencing.
\end{definition}
Note that these definitions are pchain-centric in the sense that an outfluence event on one pchain is related (by influence) to an influence event.  Since we have previously referred to events generically as influence events (and sometimes continue to do so), when it is relevant, we will clarify the nature of the event by writing `\emph{incoming influence event}' to specify that it is an influence event as defined explicitly above.

\begin{figure}
\centering
\makebox{\includegraphics[width=0.8\columnwidth]{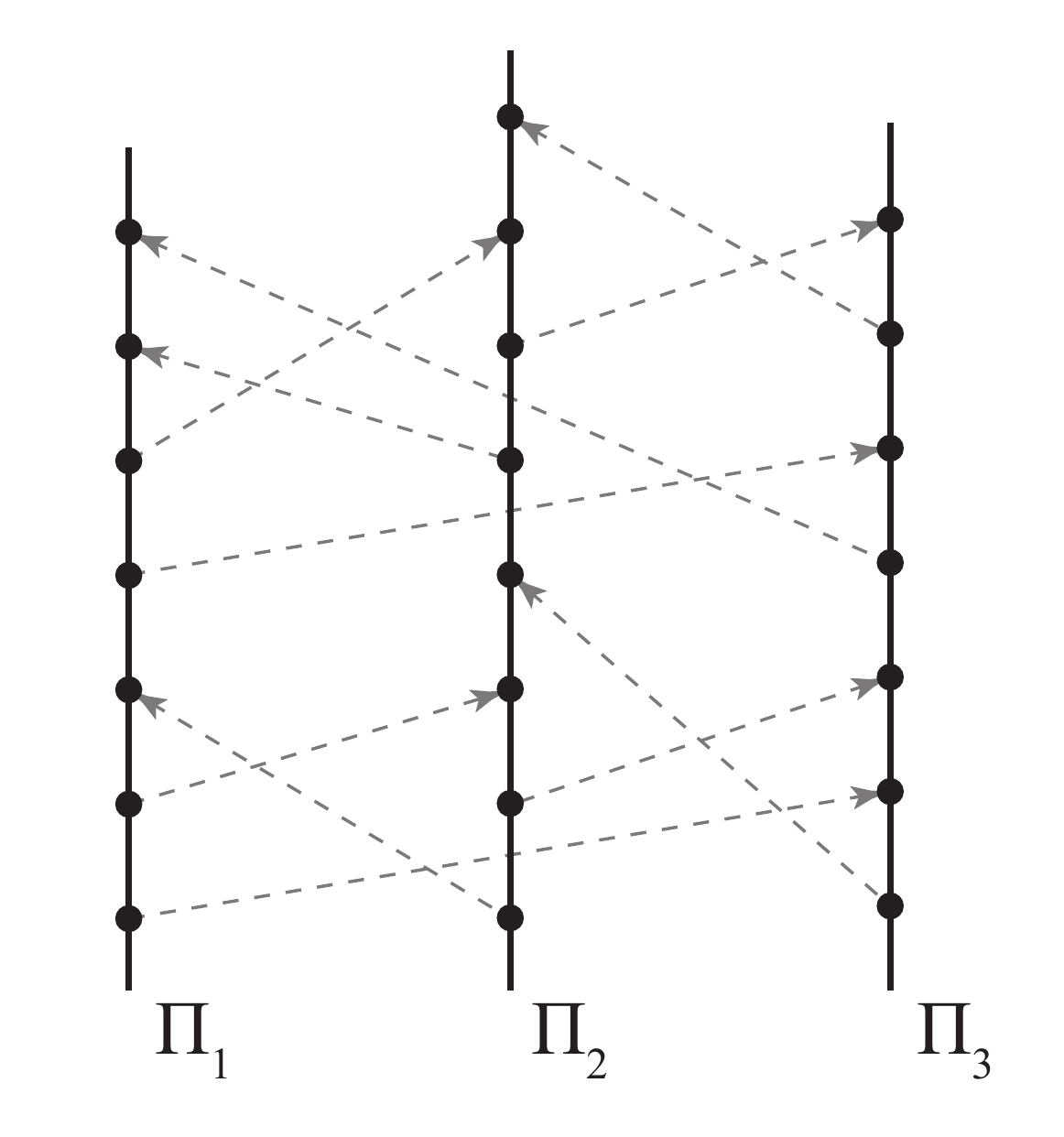}}
\caption{This figure illustrates a universe $U$ consisting of three pchains $\Pi_1$, $\Pi_2$, and $\Pi_3$.  Note that each outfluence event is related by influence (dashed lines) to an influence event, so that events come in pairs.} \label{fig:pchains}
\end{figure}

We further define a \emph{universe} of discourse (see Figure \ref{fig:pchains}) as a set of mutually influencing particles (pchains).
\begin{definition}
A \textbf{universe of discourse (universe)} $\U$ is a partially ordered set that consists of a set of mutually-influencing pchains, such that every event $p \in \U$ belongs to a pchain, $p \in \Pi \subseteq \U$.
\end{definition}

Figure \ref{fig:pchains} illustrates a small universe consisting of a set of three pchains modeling three interacting particles.  The universe is \emph{closed} in the sense that any particle influencing a particle in the universe, or influenced by a particle in the universe, is necessarily included in the universe.

Particles, each of which is modeled as a finite pchain, are distinguished from other chains composed of at least two events due to an act of influence
(illustrated with dashed lines in Figure \ref{fig:pchains}).  Traditionally, forces have been modeled by the exchange of particles called mediators.  However, since influence is not modeled by a pchain, and since pchains represent particles, this theory does not model interactions by the exchange of particles.  There is a potential advantage to this since pchains (particles) exhibit several properties reserved for fermions \cite{Knuth:FQXI2013, Knuth:Info-Based:2014}, such as Zitterbewegung and helicity in 1+1 dimensions (as will be described in Section \ref{sec:velocity}), whereas mediators are typically bosons.  As a consequence, this theory predicts an absence of supersymmetry. \footnote{The authors thank Ariel Caticha for this observation.}

The definition of a \emph{pchain}, used to model particles, and the definition of a \emph{universe} limit the structure of the posets that we will consider in this theory. \footnote{A reader preferring a different foundational model is encouraged to develop their own theory and explore its consequences.}  This restriction to a particular class of posets is an important feature that distinguishes this theory from causal set theory, which also models events as partially ordered sets, called causets, for which the ordering relation is taken to be causality. \cite{Bombelli-etal-causal-set:1987, Bombelli-etal-origin-lorentz:1989, Sorkin:2003, Sorkin:2006}  In causal set theory, it is a typical practice to randomly sprinkle events into the causet to ensure that the resulting structure is consistent with a Lorentzian manifold of the appropriate dimension.  The resulting causet is then used as a framework for discrete spacetime. \cite{Henson-causal-set-gravity:2006, Rideout+Wallden:2009}  Influence theory differs significantly from this in that the posets are not used directly to model a discrete spacetime (even though they could do so).  Instead, influence theory focuses on deriving physical laws by considering constraints on quantification imposed by symmetries.  This is the approach taken in this, and previous \cite{Knuth:FQXI2013, Knuth:Info-Based:2014, Knuth+Bahreyni:JMP2014, Walsh+Knuth:acceleration, Knuth:FQXI2015, Walsh+Knuth:event_sequences, Walsh+Knuth:geodesic, Knuth:2018, Skilling+Knuth:MPQ}, papers.

\section{Observer Chains and Consistent Quantification}
Unlike a lattice \cite{Birkhoff:1967, Gratzer2003, Davey&Priestley}, which is a poset for which each pair of elements has a unique least upper bound, the poset of events does not, in general, possess sufficient symmetry to constrain quantification \cite{Knuth:2018}. Instead, we rely on the concept of a \emph{distinguished chain} to constrain quantification of a subset of the poset.
Observer pchains will play the role of distinguished chains in that they will be independently quantified, and their quantification will constrain quantification of other events in the poset.

In this section, we review the consistent quantification of chains \cite{Knuth+Bahreyni:JMP2014}.  These results, which hold for all chains, hold for the special cases in which the chains considered are pchains.

\subsection{Quantifying Observer Chains}
We consider an observer as one who has the ability to track one or more particles and obtain information about the influence events that they experience.  In this way, the observer obtains information about events through the particles that they observe by means of influence events.  To keep the terminology concise, we will refer to an \emph{observing particle's pchain} simply as an \emph{observer chain}, or an \emph{observer}, for short.

\begin{figure}
\centering
\makebox{\includegraphics[width=0.4\columnwidth]{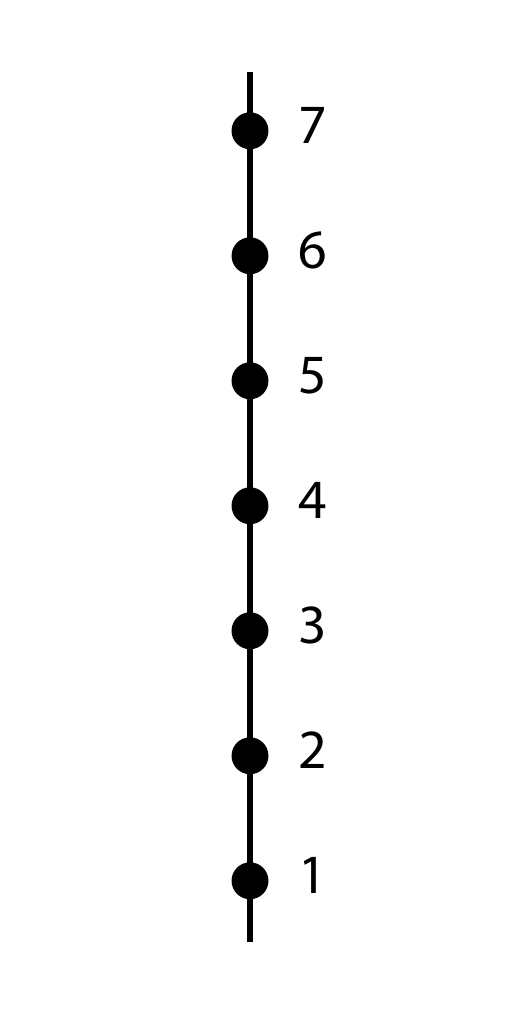}}
\caption{This figure illustrates a chain quantified by natural numbers.  All observer pchains, necessarily embedded within a universe of other pchains (not shown), are quantified similarly.} \label{fig:observer-chain}
\end{figure}

The purpose of quantification is to map events to one or more real numbers such that the numbers encode the relevant relationships among the events.  For a totally ordered set of events on a pchain one requires only a single real number per event to capture the ordering of the events.  While any valuation $v$ will do, for which $v:\mbox{event} \; x \rightarrow \mathbb{R}$ such that for all $x$, $y \in \Pi$ and for all $\Pi \in \U$ we have that $v(y) \geq v(x)$ for all $y \geq x$, the natural numbers suffice to count the events.\footnote{Nevertheless, one may be required to use real numbers to ensure consistency among a set of chains.}
A quantified chain is illustrated in Figure \ref{fig:observer-chain}.

\subsection{Chain Projection}
With the observer chains quantified by employing natural numbers to count events, it remains to quantify other events in the universe $\U$ based on their relationships to one or more observer chains.
Here we review results that hold for chains in general \cite{Knuth+Bahreyni:JMP2014}.
The critical observation is that, in general, there exist four possible classes of relationships between any event $x$ and a chain of events $\P$.
These four classes are based on whether the event $x$ includes any events on the chain $\P$, and whether the event $x$ is included by any events on the chain $\P$.  Each of these two possibilities can either be satisfied or not satisfied, which leads to four distinct relationships.

\subsubsection{Case I: No Relation}
The event $x$ has no relation to the chain, such that
there exists no event $p_x$ on the chain $\P$ that includes $x$
$$\nexists \, p_x \in \P : p_x > x,$$
and there exists no event $\overline{p}_x$ on the chain $\P$ that is included by the event $x$,
$$\nexists \, \overline{p}_x \in \P : \overline{p}_x < x.$$
This situation is illustrated in Figure \ref{fig:chain-projection}A.  In this case, the event $x$ cannot be quantified by the chain $\P$.

\subsubsection{Case II: Projection}
There exists no event $\overline{p}_x$ on the chain $\P$ that is included by the event $x$,
$$\nexists \, \overline{p}_x \in \P : \overline{p}_x < x.$$
However, there exists at least one event $p_x$ on the observer chain $\P$ that includes $x$.
Since we are assuming that the observer chains are finite, this means that there must exist a least event on the chain $\P$ that includes $x$, so that
$$\exists \, p_x \in \P : p_x > x \mbox{ and } p \ngtr x \; \forall \; p < p_x \in \P.$$
In this case, we say that the event $x$ \emph{\textit{projects}} to the chain $\P$ so that the projection of $x$ onto $\P$ is $p_x$.  When the projection $p_x$ of $x$ onto the chain $\P$ exists, it allows us to define a projection operator $P$, typically named after the chain, that operates on the event $x$ to return its projection onto the chain $p_x$, so that we can write $Px = p_x$.
This situation is illustrated in Figure \ref{fig:chain-projection}B.  Given that the event $x$ projects to the event $p_x$ on $\P$, the event $x$ can be quantified by the value $p_x$ assigned to $p_x$, by writing its coordinates based on $\P$ as $(\cdot, p_x)$.

\subsubsection{Case III: Back-Projection}
There exists no event $p_x$ on the chain $\P$ that includes the event $x$,
$$\nexists \, p_x \in \P : p_x > x.$$
However, there exists a greatest event $\overline{p}_x$ on the observer chain that is included by the event $x$
$$\exists \, \overline{p}_x \in \P : \overline{p}_x < x \mbox{ and } p \nless x \; \forall \; p > \overline{p}_x \in \P.$$
In this case, we say that the event $x$ \emph{\textit{back-projects}} to the event $\overline{p}_x$ on the chain $\P$.  When the back-projection $\overline{p}_x$ of $x$ onto the chain $\P$ exists, it allows us to define a back-projection operator $\overline{P}$, typically named after the chain, that operates on the event $x$ to return its back-projection $\overline{p}_x$ onto the chain, so that we can write $\overline{P}x = \overline{p}_x$.
This situation is illustrated in Figure \ref{fig:chain-projection}C.  Given that the event $x$ back-projects to the event $\overline{p}_x$ on $\P$, the event $x$ can be quantified by the value $\overline{p}_x$ assigned to $\overline{p}_x$, by writing $(\overline{p}_x, \cdot)$.

\subsubsection{Case IV: Both Projection and Back-Projection}
\begin{figure}
\centering
\makebox{\includegraphics[width=0.75\columnwidth]{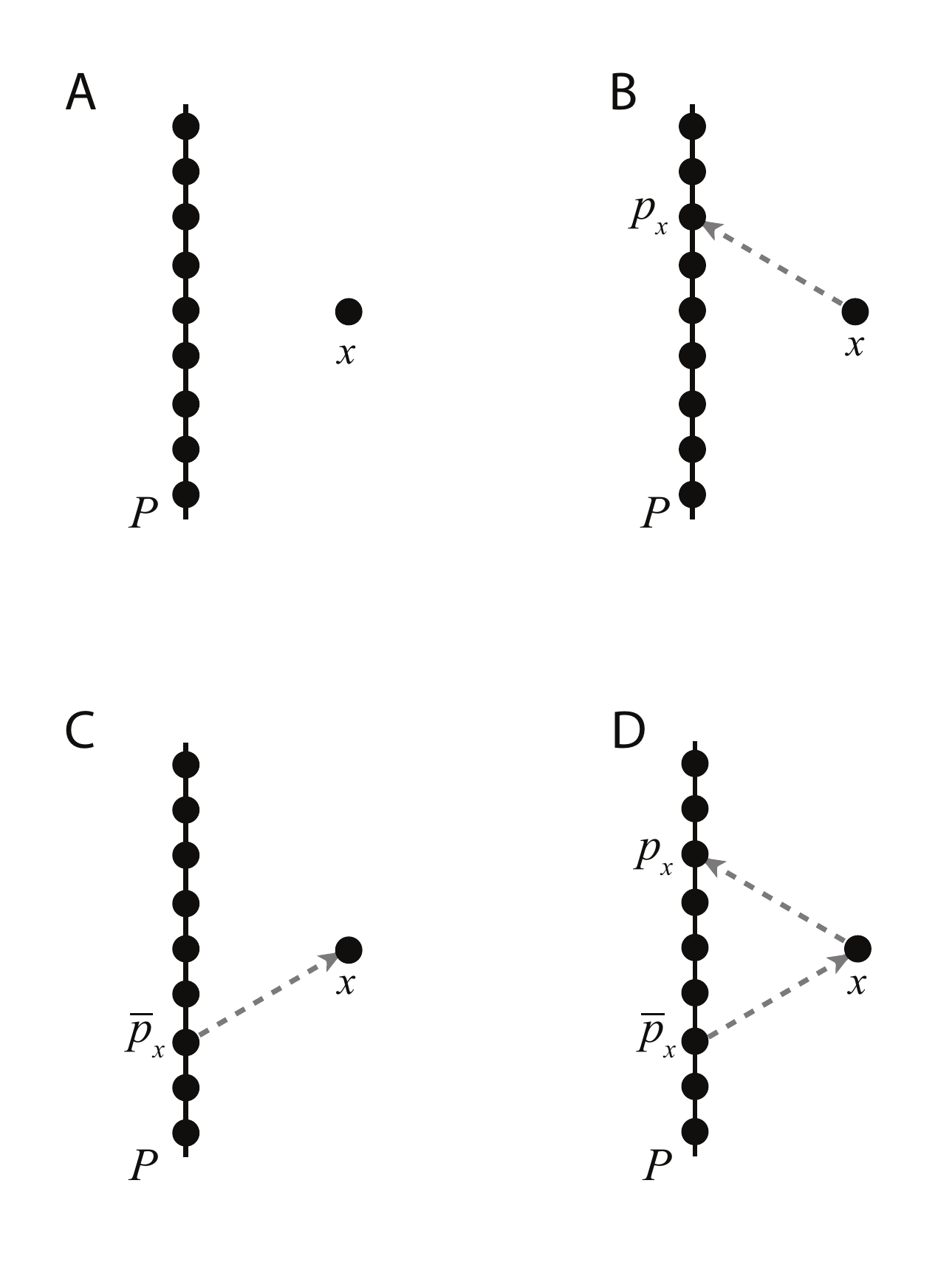}}
\caption{This figure illustrates the four possible ways in which an event can relate to a chain.  A. The event $x$ is not related to the chain $\P$.  B. The event $x$ is related to the chain $\P$ such that $x$ projects to the chain $\P$ with $p_x = Px$.  C. The event $x$ is related to the chain $\P$ such that $x$ back-projects to the chain $\P$ with $\overline{p}_x = \overline{P}x$. D. The event $x$ is related to the chain $\P$ such that $x$ both projects and back-projects to $\P$ with $p_x = Px$ and $\overline{p}_x = \overline{P}x$.} \label{fig:chain-projection}
\end{figure}

In this last case, the event $x$ both projects to an event $p_x \in \P$ and back-projects to an event $\overline{p}_x \in \P$, as illustrated in Figure \ref{fig:chain-projection}D.  Since both the projection and back-projection of $x$ onto $\P$ exist, the event $x$ can be quantified by a pair of numbers $(\overline{p}_x, p_x)$.

These relationships can be used to assign up to two unique chain-based numbers $(\overline{p}_x, p_x)$ to an event, creating an unusual observer-dependent coordinate system.  Of course, there will be events that are not related to the observer chain (Figure \ref{fig:chain-projection}A), which cannot be quantified, or events that are related in only one of the two ways, as in Figures \ref{fig:chain-projection}B and C, resulting in a single number quantification by either $(\overline{p}_x, \cdot)$ or $(\cdot, p_x)$.  In the next section, we review how these projections can be used to quantify intervals relating a pair of events.

\subsection{Coordinated Observer Chains}
In this work, we consider quantification by a pair of
observer pchains $\P$ and $\Q$.
These observer chains can be used to define a 1+1 dimensional subspace.
The subspace is defined by projecting events onto the observer chains.
An event $x$ is an element of the subspace defined by the chains $\P$ and $\Q$ if its back-projection onto one chain is consistent with its forward projection onto the other.
More specifically, an event $x$ is said to be properly collinear with the chains $\P$ and $\Q$ if one of the following cases holds \cite{Knuth+Bahreyni:JMP2014}:
\begin{equation} \label{eq:collinearity_cases}
\begin{array}{c@{}c}
    \begin{array}{ccl}
        Px = \dual{P}Qx & \qquad Qx = QPx \\
        \dual{P}x = P\dual{Q}x & \qquad \dual{Q}x = \dual{Q} \, \dual{P}x
    \end{array} & \qquad \mbox{(Case I)} \\
    \\
    \begin{array}{ccl}
        Px = P\dual{Q}x & \qquad Qx = Q\dual{P}x \\
        \dual{P}x = \dual{P}Qx & \qquad \dual{Q}x = \dual{Q}Px
    \end{array} & \qquad \mbox{(Case II)} \\
    \\
    \begin{array}{ccl}
        Px = PQx & \qquad Qx = \dual{Q}Px \\
        \dual{P}x = \dual{P} \, \dual{Q}x & \qquad \dual{Q}x = Q\dual{P}x
    \end{array} & \qquad \mbox{(Case III)} \\
\end{array}
\end{equation}
For the most part,
in this work,
we will be considering situations in which the events are collinear with the observer chains $\P$ and $\Q$.

\begin{figure}
\centering
\makebox{\includegraphics[width=0.75\columnwidth]{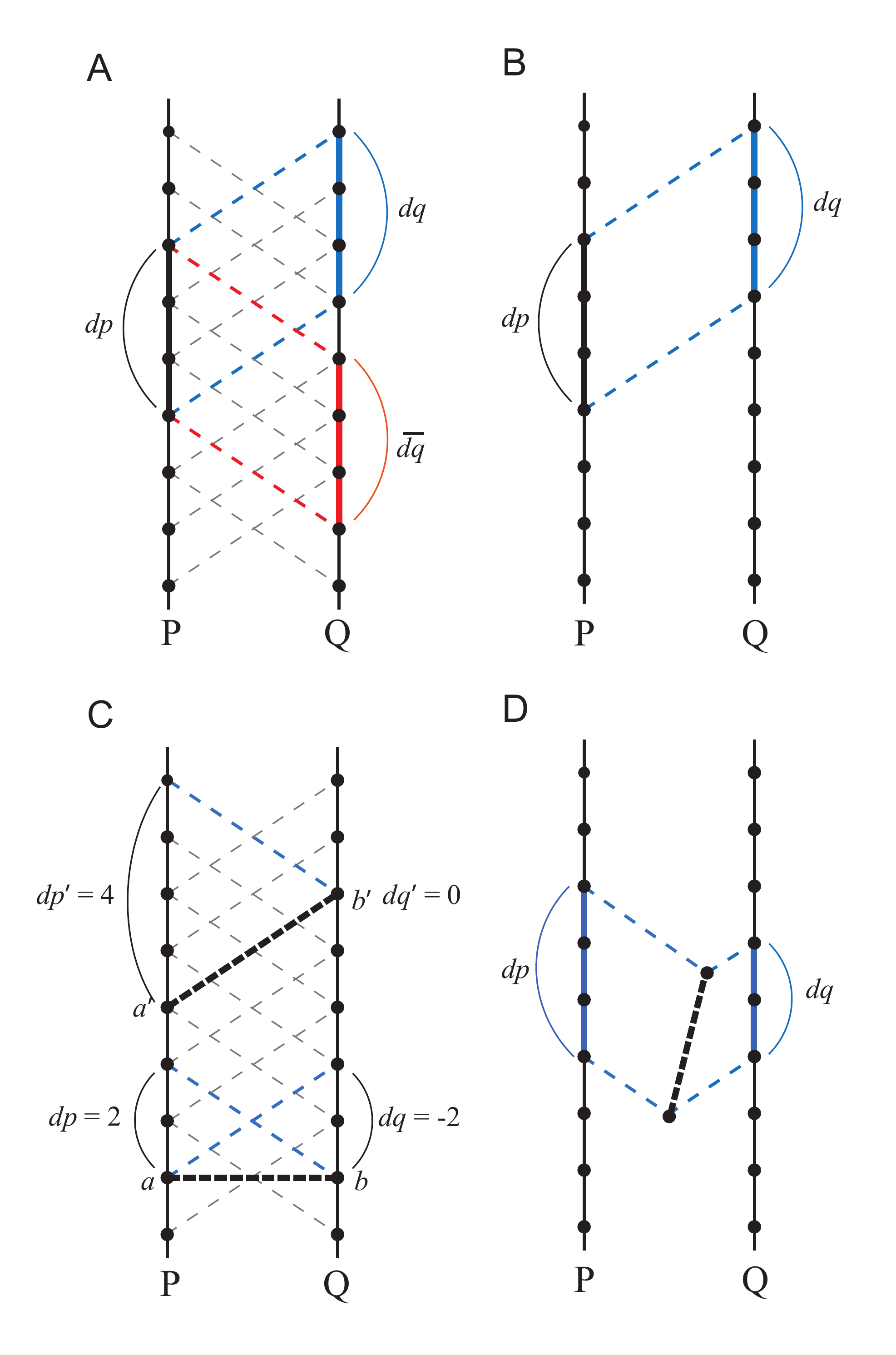}}
\caption{This figure illustrates the quantification of intervals by forward projection onto a pair of coordinated chains $\P$ and $\Q$.  A. Coordination implies that the two observer chains agree on the lengths of each other's intervals so that $\p = \q = d \! \overline{q}$.  The dashed influences represent the covering relationship between pairs of events related by an act of influence.  Coordination allows observers to quantify intervals with only forward projections.  B.  Intervals along chains represent time-like intervals and are quantified by the average of the projected lengths onto the observer chains (\ref{eq:dt}).  C. The relationship between chains is quantified uniquely by computing the half difference of the projections (\ref{eq:dx}).  For coordinated observers, this relationship, called directed distance, does not depend on which events were considered.  This is illustrated by two cases in which $dx = (dp - dq)/2 = (2 - (-2))/2 = 2$ and $dx = (dp' - dq')/2 = (4 - 0)/2 = 2$.  D. Generalized intervals defined by any pair of events are uniquely quantified by the pair $(dp,dq)$ and the
squared length $ds^2 = {dp}\,{dq}$.}
\label{fig:influence-examples}
\end{figure}

Moreover, we assume that the observer chains exhibit a relationship called coordination \cite{Knuth+Bahreyni:JMP2014}.  Basically, coordination implies that the two observer chains agree on the
lengths of each other's intervals (Figure \ref{fig:influence-examples}A).
That is, every interval on $\P$ projects to an interval on $\Q$ with the same length, and vice versa.  The coordination condition is essentially equivalent to the synchronization of co-moving clocks, which will result in a flat space-time.\footnote{The coordination condition can be relaxed.}  As a result, back-projections onto one chain are equivalent to forward projections onto the other chain.  This means that the length of an interval can be formulated entirely in terms of forward projections onto the two observer chains.

\section{Quantification of Intervals}
In this section, we review the quantification of intervals by two coordinated observer chains $\P$ and $\Q$.  More details, including proofs of uniqueness, can be found in Knuth and Bahreyni \cite{Knuth+Bahreyni:JMP2014}.

Consider an interval $I = [a, b]$, defined by two events $a$ and $b$, that forward projects to both $[p_a, p_b] \in \P$ and $[q_a, q_b] \in \Q$.  The interval $I$ can then be quantified by the 4-tuple
$$(p_a, p_b, q_a, q_b).$$
We can also consider the lengths $dp = p_b - p_a$ and $dq = q_b - q_a$ of the intervals to which the interval $I$ projects, and quantify the interval $I$ with the pair of lengths
$$(dp, dq) = (p_b - p_a, q_b - q_a).$$

Last, we have demonstrated \cite{Knuth+Bahreyni:JMP2014} that a unique scalar can be associated with each interval based on the product of its projections onto the pair of coordinated observer chains (Figure \ref{fig:influence-examples}D).  That is, there exists a function $\sigma^2$ that takes a pair of projected lengths and returns a unique scalar that is a squared length
\begin{equation} \label{eq:ds2}
ds^2 = \sigma^2(dp, dq) = {dp}\,{dq}.
\end{equation}
This results from the fact that the observer chains can be quantified with any fixed increment.  If we write the function that takes the projected lengths $dp$ and $dq$ to the length of the interval as $\sigma(dp, dq)$, then allowing for a regraduation (rescaling) of the quantification of the observer chains by some scale $\alpha$ that also rescales the scalar assigned to the interval by the same factor $\alpha$ we have
\begin{equation}
\alpha \sigma(dp, dq) = \sigma(\alpha dp, \alpha dq).
\end{equation}
This implies that the function $\sigma$ must be of the form
\begin{equation}
\sigma(dp, dq) = \sqrt{dp \, dq} \;\; h\left(\frac{dp}{dq}\right),
\end{equation}
where $h\left(\frac{dp}{dq}\right) = h\left(\frac{dq}{dp}\right)$.  Since either $dp$ or $dq$ can be negative, it is better to work with the function $\sigma^2$ and write
\begin{equation}
\sigma^2(dp, dq) = {dp} \, {dq} \;\; h^2\left(\frac{dp}{dq}\right).
\end{equation}
Requiring that assigned lengths be invariant under linear transformations results in $h$ being identically one, so that
\begin{equation}
\sigma^2(dp, dq) = {dp} \, {dq}.
\end{equation}

We have also demonstrated that with two coordinated observer chains there are two distinct measures of length \cite{Knuth+Bahreyni:JMP2014}.  The first measure of length concerns intervals that are defined along chains
\begin{equation} \label{eq:dt}
dt = \frac{1}{2}(dp + dq),
\end{equation}
so that it is associated with a concept of time (Figure \ref{fig:influence-examples}B).
The second measure quantifies the relationship between chains
\begin{equation} \label{eq:dx}
dx = \frac{1}{2} (dp - dq),
\end{equation}
so that it is associated with a concept of directed distance (Figure \ref{fig:influence-examples}C).
We refer to (\ref{eq:dt}) and (\ref{eq:dx}) as the space-time decomposition \cite{Knuth+Bahreyni:JMP2014}.

Pair quantifications of intervals, which can be written in terms of $dt$ and $dx$ by
\begin{equation}
(dp, dq) = (dt, dt) + (dx, -dx),
\end{equation}
are additive.
This is reminiscent of Bondi's k-calculus \cite{Bondi:1980} and Kauffman's iterant algebra \cite{Kauffman:1985}.
The scalar quantification is also additive resulting in
\begin{align}
ds^2 &= \sigma^2 \Big( dp, dq \Big) = {dp}\,{dq} \nonumber \\
&= \sigma^2 \Big( (dt, dt) + (dx,-dx) \Big) \nonumber \\
&= \sigma^2 \Big( (dt, dt) \Big) + \sigma^2 \Big( (dx,-dx) \Big) \nonumber \\
&= dt^2 - dx^2,
\end{align}
which is the Minkowski metric \cite{Knuth+Bahreyni:JMP2014}.
It should be noted that this result is, in part, a consequence of the assumption that the observer chains are coordinated.  This means that they agree on the lengths of each other's intervals, which is equivalent to the traditional situation in which their clocks are synchronized.  Together, coordinated observer chains define an inertial frame.

\section{Kinematics}
In classical mechanics, the term kinematics refers to the description of the motion of objects without considering the causes of such motion.  In influence theory, the term \emph{kinematics} can be thought of as referring to the description of an object's influence on others.  This use of the term will be demonstrated to be appropriate, since the descriptions of an object's influence on others result in descriptions of an object's change in position as a function of time, and its momentum and energy.  To focus on an object's influence on others, we begin by defining the idealization of a free particle.

\subsection{The Free Particle}
Influence theory models an elementary object, such as a particle, as a totally ordered finite pchain of influence events.  In general, these events represent both the particle influencing others, and the particle being influenced by others.  The goal is to develop a general theory to describe and predict the behavior of objects based on what is known about these influence events from the events experienced by the observer chains.  A general theory applies to all cases, and our strategy \cite{Skilling+Knuth:MPQ} has been to apply \emph{eliminative induction} \cite{Caticha2012:entropic} to eliminate theories that give the wrong results in simple situations leaving a
single candidate theory that then, if a general theory exists, must be the unique general theory.

Idealizations are situations where the expected results are clear, and as such, they make excellent cases for eliminative induction.  We consider two idealizations, which will be useful in thought-experiments.  At one extreme, a \emph{free particle} is defined as an object that influences others with constant rates of influence, but is not itself influenced.  At the other extreme we define a \emph{quenched particle} as an object that is influenced by others, but does not influence others.  Real particles, encountered in the laboratory, are expected to lie somewhere within the spectrum defined by these two extremes.

One might be tempted to define a free particle as a particle that neither influences nor is influenced.  If this were the case, then the particle would not be part of the universe (by definition above) since it would not interact with any particles in the universe.  No observer could possibly know about it.
We have already demonstrated that time intervals and lengths are quantities assigned by observers based on the influences that they receive.  For this reason, a free particle is assumed to influence others while not being influenced by others.  This is necessary so that the observers can know about the free particle and use familiar quantities (time and position) to describe it.  However, the main point is that the free particle is not influenced by others.  We shall show that when a particle is influenced, such influence plays the role of a force.  The un-influenced free particle is essentially free of outside forces.  At the other extreme, the quenched particle is influenced by others, but does not influence others.  While there are interesting things to say about quenched particles, we will save a thorough discussion for a future work.

We begin by considering a universe consisting of a free particle $\PF$ that influences two coordinated observer chains $\PO$ and $\QO$, such that the particle chain $\PF$
is situated between
$\PO$ and $\QO$, with $\PO < \PF < \QO$.  The observer chains $\PO\QO$ define an inertial frame for the particle \cite{Knuth+Bahreyni:JMP2014}.
Since the free particle $\PF$ can only influence the two observer chains, the only information available about the behavior of $\PF$ is that for each event experienced by $\PF$, either $\PO$ was influenced or $\QO$ was influenced.  As a result, one can only count the two kinds of events. Thus one should expect that the resulting theory for the free particle, at its core, should be based on event counting.

\subsection{Event Counting}
With event counting, it would be reasonable to assign values of $\ddp = 1$ and $\ddq = 1$  to the lengths of each interval defined by successive influence events on observer chains $\PO$ and $\QO$, respectively.
This amounts to a choice of units, and as such, it does not limit the generality of the approach for the single particle.  However, with the expectation that a multi-particle theory might allow for intervals of different lengths, we quantify the length of the interval between successive events on the particle chain with a basic unit written as $\delta_\tau$.  For an interval on the particle chain defined by an extremely large number $N$ of successive events, the interval length, or \emph{proper time}, is given by
\begin{equation}
\dtau = (N-1) \delta_\tau \approx N \delta_\tau. \label{eq:N-delta-tau}
\end{equation}
It is useful to first consider the case in which the observer chains are coordinated with the particle chain, such that an interval of length $\dtau$ containing a large number $N$ of events on the particle chain projects to intervals of equal lengths on the observer chains $\PO$ and $\QO$ so that
\begin{equation}
\dtau = \dpo = \dqo. \label{eq:dp&dq}
\end{equation}
We will demonstrate that these assumptions lead to the fact that the frame defined by $\PO\QO$ acts as a rest frame.

The assumption $\dpo = \dqo$ is consistent with Laplace's Principle of Indifference \cite{Keynes:2013}, which assumes that there is no reason for the free particle $\PF$ to prefer to influence one observer chain over the other so that the probability, $\PrPO$, that $\PF$ will influence $\PO$ is equal to the probability, $\PrQO$, that $\PF$ will influence $\QO$:
\begin{equation}
\PrPO = \PrQO = \frac{1}{2}.
\end{equation}
For $N$ events experienced by the free particle $\PF$,
\begin{equation}
\Np = N \PrPO = \frac{N}{2}
\end{equation}
events will influence the observer chain $\PO$,
and
\begin{equation}
\Nq = N \PrQO = \frac{N}{2}
\end{equation}
events will influence the observer chain $\QO$.

Since event counting implies that $\ddp = \ddq = 1$,
\begin{equation}
\dpo = \Np \ddp = \frac{N}{2}.
\end{equation}
These $\Np$ influence events result in an interval of projected length $\frac{N}{2}$ on $\PO$.
However, since the particle chain is coordinated with the observer chains, we have that
\begin{equation} \nonumber
\dtau = \dpo,
\end{equation}
which implies that
\begin{equation}  \nonumber
N \ddtau = \frac{N}{2},
\end{equation}
so that
\begin{equation}
\ddtau = \frac{1}{2}.
\end{equation}
In short, we have that
\begin{align}
\dpo &= \dtau = \Np = N \ddtau = \frac{N}{2}  \label{eq:dpo}\\
\dqo &= \dtau = \Nq = N \ddtau = \frac{N}{2}. \label{eq:dqo}
\end{align}

In this 1+1-dimensional single-particle theory, the units chosen by assigning unity to the length of an interval between successive events on the observer chains force the fundamental unit of length $\ddtau$ on the particle chain to be $\ddtau = \frac{1}{2}$.
This is consistent with the more general relationship
\begin{equation}
\dtau = \sqrt{\p \q}
\end{equation}
for which
\begin{equation}
N \ddtau = \dtau = \sqrt{\dpo \dqo} = \sqrt{\frac{N^2}{4}} = \frac{N}{2}.
\end{equation}
As a result, the fundamental length scale assigned to the particle is $\frac{1}{2}$ so that half-integers are seen to play a natural role in the theory.

In general, the fundamental length scales are expected to be extremely small (by at least a factor of $10^{-21}$) relative to the interval lengths at macroscopic scales, which correspond to extremely large $N$.  Since in this paper we are more interested in macroscopic scales, we write these extremely small quantities as differentials.

Note that since $\PF$ either influences $\PO$ or $\QO$, the projected events experienced by the observers cannot be ordered with respect to one another.
As a result, no observer can possibly know the sequence of events experienced by the particle $\PF$.  This is an example of information isolation, which is a critical feature of quantum mechanics \cite{Schumacher+Westmoreland:2010}.
As a result, definitive space-time coordinates cannot be assigned to any event experienced by this free particle.  This has important implications at the microscopic (quantum) scale of individual events. Since it is not possible to define space-time positions for a free particle, one cannot describe the behavior of a free particle in terms of a space-time path or trajectory.  Furthermore, one can show that to make inferences about a free particle's behavior, one must consider the set of all possible orderings that are consistent with the observations \cite{Knuth:Info-Based:2014},
which is equivalent to considering the set of all possible paths that the particle could have taken.
This interesting aspect of the theory will be discussed in a forthcoming paper focused on influence theory and quantum mechanics.

\subsection{Velocity of the Free Particle} \label{sec:velocity}
The \emph{average velocity}, $\beta$, can be defined \cite{Knuth:Info-Based:2014} over an extremely large number $N$ of events with respect to a pair of coordinated observers $\PO$ and $\QO$ from (\ref{eq:dt}) and (\ref{eq:dx}) as
\begin{equation}
\beta_{\PO\QO} = \frac{\dxo}{\dto} = \frac{\dpo - \dqo}{\dpo + \dqo}.
\end{equation}
Note that the magnitude of the velocity has a maximum at $|\beta| = 1$ obtained by
setting either $\dpo = 0$ or $\dqo = 0$.
This implies that the theory naturally imposes a finite ultimate speed that cannot be exceeded.
Moreover, since we are considering the case $\dpo = d \! q_0$, the average velocity is identically zero so that the frame $\PO\QO$ is recognized as being the \emph{rest frame} of the particle.

\begin{figure}
\centering
\makebox{\includegraphics[width=0.75\columnwidth]{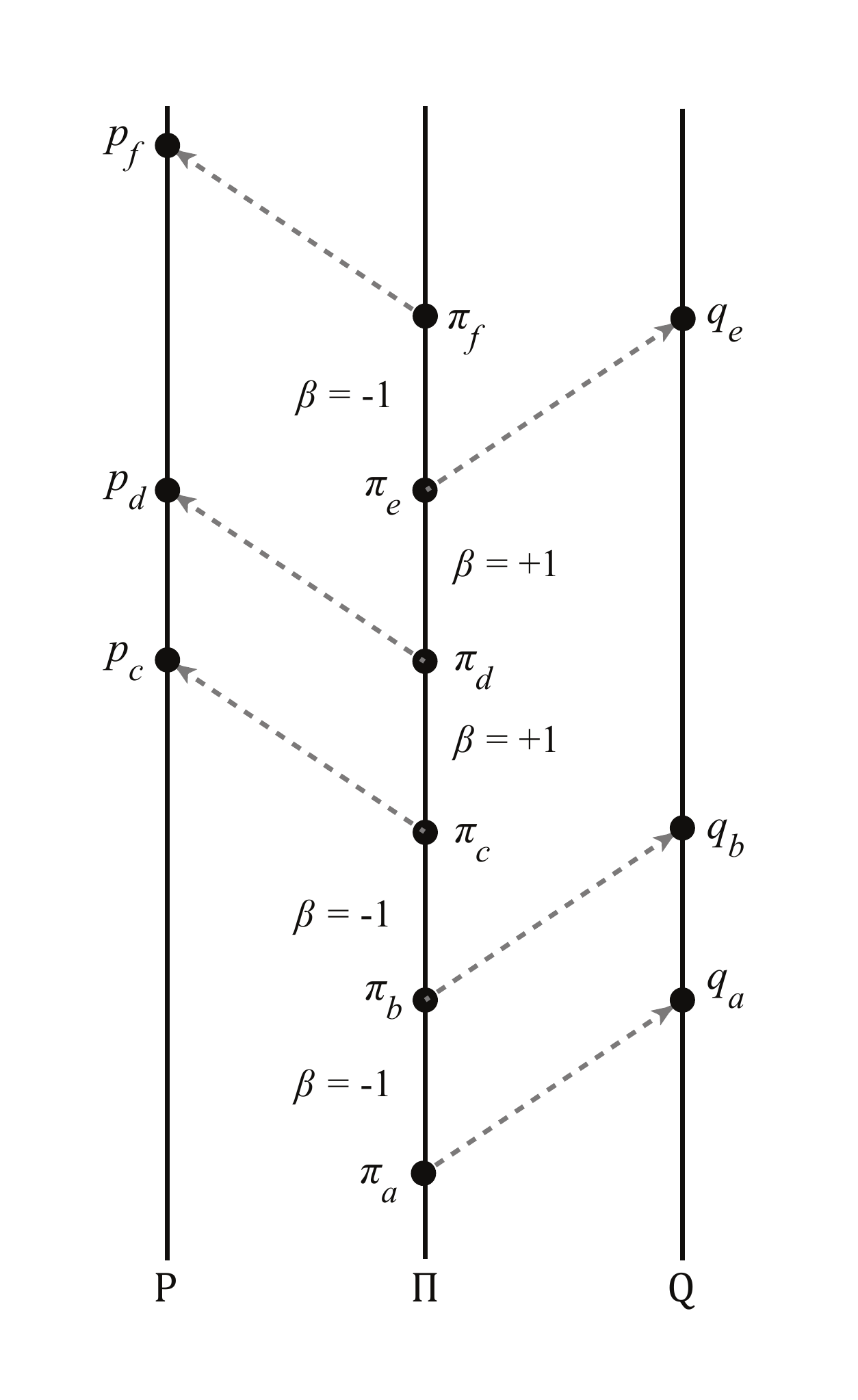}}
\caption{A. This figure illustrates how the free particle exhibits \emph{Zitterbewegung} for which at the microscopic scale the particle zig-zags back and forth at the ultimate speed. Each atomic interval along the particle chain $\Pi$ projects onto the observer chains $\P$ and $\Q$ with either $\p = 0$ or $\q = 0$ so that for each interval $\beta = \pm 1$. For example, the interval $[\pi_b, \pi_c]$ projects to the interval $[q_b, q_e]$ with $\q \neq 0$ on $\Q$ and to the degenerate interval $[p_c, p_c]$ with $\p = 0$ on $\P$ so that $\beta = -1$.} \label{fig:zitter}
\end{figure}

Since average velocities are associated with intervals, there is no concept of an instantaneous velocity associated with a single event in this theory.  The closest one can get is to consider the average velocity associated with an \emph{atomic interval} on the particle chain defined by successive influence events $(\pi_a \prec \pi_b) \in \PF$.  There are two relevant classes of atomic interval (Fig. \ref{fig:zitter}) based on whether the least event, $\pi_a$, represents the fact that the free particle influenced $\P$ or whether it represents the fact that the free particle influenced $\Q$.
The case of $\pi_a$, in Figure \ref{fig:zitter}, is the latter case, since $\pi_a$ influences $\Q$ so that $Q \pi_a \neq Q \pi_b$, giving $\q \neq 0$ and $\p = 0$ because both $\pi_a$ and $\pi_b$ project to the same event, $p_c$ in $\P$.
In the former case, exemplified by the interval $[\pi_c, \pi_d]$, in Figure \ref{fig:zitter}, we have that $\q = 0$ and $\p \neq 0$.
As a result, the velocity $\beta_p$ assigned to atomic interval $[\pi_c, \pi_d]$, for which the
least event,
$\pi_c$, influences $\P$, is characterized by $\q = 0$ so that
\begin{equation}
\beta_p = \frac{\p - 0}{\p + 0} = +1.
\end{equation}
Similarly, for an atomic interval, such as $[\pi_a, \pi_b]$, for which the
least event,
$\pi_a$, influences $\Q$, the velocity $\beta_q$ assigned to that interval is
\begin{equation}
\beta_q = \frac{0 - \q}{0 + \q} = -1.
\end{equation}
As a result, at the microscopic level, the free particle $\PF$ zig-zags back and forth at the ultimate speed in a motion known as \emph{Zitterbewegung}, which is a quantum effect predicted by the Dirac equation \cite{Breit:1928, Schrodinger:1930}.
This suggests that the outfluence events of the free particle and its associated zitter could be interpreted as the particle interacting with a Higgs field \cite{Hestenes:2008electron}.
There is no spin in this 1+1-dimensional description of a free particle.  However, one can define the helicity based on whether the particle last influenced $\P$ or $\Q$.
Given $N$ consecutive events for which $N = \Np + \Nq$ and $\Np = \Nq$, the average velocity, which is the average of these microscopic velocities, can be shown to be
\begin{align}
\beta &= \PrPO - \PrQO \\
&= \frac{\Np - \Nq}{N} = 0.
\end{align}
As a result, the free particle can be at rest only on average.  At microscopic scales, the particle is always moving at the ultimate speed.  Thus, in this theory, there is no such thing as a rest frame at all scales.  This suggests that general relativity ought to be on shaky ground at microscopic (quantum) scales.

\subsection{Linearly Related Inertial Frames}

For greater generality, we consider the case in which a pair of observers $\P$ and $\Q$ are linearly related to the pair of observers $\PO$ and $\QO$ defining the rest frame of the particle.  In this case, the pair $(\dpo, \dqo)_{\PO\QO}$, used to quantify an interval with respect to $\PO$ and $\QO$, is related to the pair $(\p ,\q)_{\P\Q}$ used to quantify the same interval by $\P$ and $\Q$ by the linear pair transformation $L_{\PO\QO\rightarrow\P\Q}$ \cite{Knuth+Bahreyni:JMP2014}:
\begin{align}
(\p , \q) &= L_{\PO\QO\rightarrow\P\Q}( (\dpo, \dqo) )\\
&= (k \dpo, \frac{1}{k} \dqo), \label{eq:pair-transformation}
\end{align}
so that
\begin{align}
\p &= k \dpo           \label{eq:dp_k}\\
\q &= \frac{\dqo}{k}.  \label{eq:dq_k}
\end{align}
%

As a result, the average velocity of the particle in any inertial frame defined by coordinated observers, and the constant $k$,
is given by
\begin{equation}
\beta = \frac{k\,\dpo - \frac{1}{k}\,\dqo}{k\,\dpo + \frac{1}{k}\,\dqo}.
\end{equation}
Since in the rest frame of the particle, $\dpo = \dqo$, the average velocity can be written succinctly as
\begin{equation}
\beta = \frac{k - \frac{1}{k}}{k + \frac{1}{k}}. \label{eq:beta}
\end{equation}
Solving for $k$ gives
\begin{equation}
k = \sqrt{\frac{1+\b}{1-\b}} = 1+z,
\end{equation}
where $z$ is the relativistic Doppler \emph{redshift}.

It will be important to compute the \emph{Lorentz factor}
\begin{align}
\gamma &= \frac{1}{\sqrt{1-\beta^2}} = \frac{dt}{d\tau}\\
&= \frac{\p + \q}{2\sqrt{\p \q}}\\
&= \frac{k+\frac{1}{k}}{2}, \label{eq:gamma-k}
\end{align}
which relates the proper time $\dtau$ experienced by the particle to the time $dt$ experienced by the observers
\begin{align}
dt &= \gamma \dtau\\
&= \frac{k+\frac{1}{k}}{2} \dtau. \label{eq:gamma}
\end{align}
In the rest frame, the parameter $k = 1$, and we have that $dt = \dtau$ so that the passage of time experienced by the particle is identical to that experienced by the observers.

The space-time coordinates in the rest frame
\begin{align}
  \dto &= \frac{1}{2}(\dpo + \dqo) \\
  \dxo &= \frac{1}{2}(\dpo - \dqo)
\end{align}
transform to
\begin{align}
  \dt &= \frac{1}{2} \left( k\dpo + \frac{1}{k}\dqo \right)  \label{eq:dt:k}\\
  \dx &= \frac{1}{2} \left( k\dpo - \frac{1}{k}\dqo \right), \label{eq:dx:k}
\end{align}
in a linearly related frame.
They can be further simplified to
\begin{align}
  \dt &= \gamma \dto + \gamma (\beta) \dxo \\
  \dx &= \gamma (\beta) \dto + \gamma \dxo,
\end{align}
or
\begin{align}
  \dt &= \gamma \dto - \gamma (-\beta) \dxo \\
  \dx &= -\gamma (-\beta) \dto + \gamma \dxo,
\end{align}
which are the standard \emph{Lorentz transformations} where $-\beta$ is the velocity of observers $\P\Q$ with respect to $\PO\QO$.  Thus influence theory is fundamentally relativistic (without making any assumptions about the speed or nature of light).

\subsection{Energy, Momentum and Mass}

While one can study events and event sequences by considering intervals, one can alternatively quantify events by considering rates of influence.
We begin by defining the rates, $\rpo$ and $\rqo$, so that
\begin{equation}
dN = \rpo \, \dpo + \rqo \, \dqo.
\end{equation}
We consider $N$ outfluence events,
which we write as $dN$ because the corresponding lengths are much smaller than any lengths of interest in the macroscopic scales we are considering.  The total number of outfluence events $dN$ is equal to the sum of the $dN_{\P}$ events
of influencing $\P$
and the $dN_{\Q}$ events
of influencing $\Q$
\begin{equation}
dN = dN_{\P} + dN_{\Q}.
\end{equation}
We now take advantage of the fact that intervals between successive events of influencing the observers
are much smaller than macroscopic scales, so that enough events to establish approximate continuity can be grouped and still constitute approximately differential amounts.  A similar approach is taken in fluid dynamics.
By writing the total differential in terms of partial differentials
\begin{equation}
dN = \frac{\partial N}{\partial p_0} \dpo + \frac{\partial N}{\partial q_0} \dqo,
\end{equation}
we can write the rates in terms of the partial derivatives
\begin{align}
\rpo &= \frac{\partial N}{\partial p_0} = dN_{\P} \; {\dpo}^{-1}  \label{eq:rpo-partial}\\
\rqo &= \frac{\partial N}{\partial q_0} = dN_{\Q} \; {\dqo}^{-1} \label{eq:rqo-partial}
\end{align}
so that rates are inverse lengths.

From (\ref{eq:dpo}) we have that
\begin{equation}
\dpo = dN \; \ddtau
\end{equation}
and that
\begin{equation}
dN = 2 \; dN_{\P},
\end{equation}
where we are writing small quantities as differentials.
This implies that
\begin{equation}
\dpo = 2 \; dN_{\P} \; \ddtau. \label{eq:dN_P}
\end{equation}
Similarly, from (\ref{eq:dqo}) we have that
\begin{equation}
\dqo = 2 \; dN_{\Q} \; \ddtau. \label{eq:dN_Q}
\end{equation}
Substituting (\ref{eq:dN_P}) into (\ref{eq:rpo-partial}), and (\ref{eq:dN_Q}) into (\ref{eq:rqo-partial}), we get
\begin{align}
\rpo &= \frac{1}{2 \ddtau} \\
\rqo &= \frac{1}{2 \ddtau}
\end{align}
so that, in general, we have from (\ref{eq:dp_k}) and (\ref{eq:dq_k}) that
\begin{align}
\rp &= \frac{1}{k}\frac{1}{2 \ddtau} \\
\rq &= k\frac{1}{2 \ddtau}.
\end{align}

We now relate the rates of influence to the familiar concepts of energy and momentum based on the fact that the following rate-based quantities exhibit the requisite properties.
The \emph{energy} of the particle is defined as the (symmetric) average, or arithmetic mean, of the rate of influence
\begin{align}
E &= \frac{1}{2}(\rp + \rq) \label{eq:energy;r}\\
&= \frac{1}{4 \delta_{\tau}} \left( k + \frac{1}{k} \right),\label{eq:energy;k}
\end{align}
which can also be written as
\begin{equation}
E = \frac{dN}{2} \frac{\dt}{\p\q} = \frac{dN}{2} \frac{\dt}{\dtau^2}. \label{eq:energy;dtau}
\end{equation}
Similarly, the \emph{momentum} of the particle is defined as the (anti-symmetric) half-difference of the rates of influence
\begin{align}
P &= \frac{1}{2}(-\rp + \rq) \label{eq:momentum;r}\\
&= \frac{1}{4 \delta_{\tau}} \left( k - \frac{1}{k} \right), \label{eq:momentum;k}
\end{align}
which can be written as
\begin{equation}
P = \frac{dN}{2} \frac{\dx}{\p\q} = \frac{dN}{2} \frac{\dx}{\dtau^2}. \label{eq:momentum;dtau}
\end{equation}
With either (\ref{eq:energy;k}) and (\ref{eq:momentum;k}), or (\ref{eq:energy;dtau}) and (\ref{eq:momentum;dtau}), it is easily shown that
\begin{equation}
\beta = \frac{P}{E},
\end{equation}
as expected.

The \emph{rest mass}, or \emph{invariant mass}, of the particle is then given by $M^2 = E^2 - P^2$ so that
\begin{align}
M &= \sqrt{r_{\P} r_{\Q}}\\
&= \frac{1}{2 \ddtau}, \label{eq:mass}
\end{align}
which in the adopted units gives $M = 1$.
As expected from quantum mechanics, the energy and momentum, as defined in terms of rates, are the Fourier duals of the time and displacement, respectively, which are defined in terms of intervals.

By using (\ref{eq:gamma-k}), (\ref{eq:energy;k}), and (\ref{eq:mass}), the energy can be written as
\begin{equation}
E = \gamma M,
\end{equation}
and by using (\ref{eq:beta}), (\ref{eq:gamma-k}), (\ref{eq:momentum;k}), and (\ref{eq:mass}), the momentum can be written as
\begin{align}
P &=  \frac{1}{4 \ddtau} \left( k - \frac{1}{k} \right) \\
&=  \frac{1}{4 \ddtau} \frac{k + \frac{1}{k}}{k + \frac{1}{k}} \left( k - \frac{1}{k} \right) \\
&= \frac{1}{2 \ddtau} \frac{k + \frac{1}{k}}{2}  \frac{k - \frac{1}{k}}{k + \frac{1}{k}} \\
&= M \cdot \gamma \cdot \beta\\
&= \gamma \beta M, \label{eq:P=gbM}
\end{align}
as expected.

The rates at which the free particle influences the observers can be written in terms of its rest mass as
\begin{align}
\rp &= \frac{1}{k} \frac{1}{2 \ddtau} = \frac{1}{k} M\\
\rq &=  k \frac{1}{2 \ddtau} = kM,
\end{align}
so that the rate at which a particle influences others is directly proportional to its rest mass.

There are two ways to conceive of mass in special relativity.  The first is the rest mass or invariant mass, which is the observer-independent magnitude of the four-momentum, and the second is the relativistic mass, which is observer-based and hence dependent on the velocity of the object.  Some insights are obtained by carefully considering the way in which rates of influence are defined for an observed free particle.  There are at least two ways in which this can be done.

The first method is to consider an interval defined by $N+1$ events on the particle chain and to consider the projection of this interval onto the observer chains.  This interval of length $\frac{N}{2}$ on the particle chain projects to intervals of length $\p = k \frac{N}{2}$ and $\q = \frac{1}{k} \frac{N}{2}$ on observer chains $\P$ and $\Q$, respectively.
One then defines rates consistent with the general expressions given in the first equalities of (\ref{eq:rpo-partial}) and (\ref{eq:rqo-partial}), noting that $k$ is constant for linearly related chains, as
\begin{align}
r_p &= \frac{dN_{\P}}{dp} = \frac{dN/2}{k \frac{dN}{2}} = \frac{1}{k}\\
r_q &= \frac{dN_{\Q}}{dq} = \frac{dN/2}{\frac{1}{k} \frac{dN}{2}} = k.
\end{align}
The mass $M$ is then given by
\begin{equation}
M = \sqrt{r_p r_q} = \sqrt{\frac{1}{k} k} = 1, \label{eq:M}
\end{equation}
which is clearly invariant.  This is the \emph{rest mass} or the \emph{invariant mass}.

The second method is to consider two intervals of equal length $\p =\q$ along the observer chains and to count the events $\Np$ and $\Nq$, with $N = \Np + \Nq$.  The total rate of influence with respect to the intervals $dp$ and $dq$ is then defined to agree with the rates above in the case that the observer intervals correspond to a single particle chain interval, which occurs in the rest frame:
\begin{align}
\tilde{r}_p &= \frac{N/2}{dp} = \frac{N_{\P} + N_{\Q}}{2 N_{\P} k} \\
\tilde{r}_q &= \frac{N/2}{dq} = \frac{N_{\P} + N_{\Q}}{2 N_{\Q} \frac{1}{k}}.
\end{align}
The mass $\tilde{M}$ is then given by
\begin{equation}
\tilde{M} = \sqrt{\tilde{r}_p \tilde{r}_q} =  \sqrt{\frac{(N_{\P} + N_{\Q})^2}{2 N_{\P} k \cdot 2 N_{\Q} \frac{1}{k}}}.
\end{equation}
To simplify this, we use the fact that we are considering intervals of equal length $\p = \q$ along  the observer chains for which $\p = \Np k$ and $\q = \Nq \frac{1}{k}$.  This implies that $\Nq = \Np k^2$.  The mass is then
\begin{align}
\tilde{M} &= \sqrt{\frac{(\Np + \Nq)^2}{2 \Np k \cdot 2 \Nq \frac{1}{k}}}\\
&= (\Np + \Nq) \sqrt{\frac{1}{2 \Np k \cdot 2 \Nq \frac{1}{k}}} \\
&= \Np (1 + k^2) \sqrt{\frac{1}{2 \Np k \cdot 2 \Np k}} \\
&= \frac{1 + k^2}{2 k} \\
&= \gamma, \label{eq:tildeM}
\end{align}
where in the last step we used (\ref{eq:gamma-k}).

Comparing the two masses $M$ (\ref{eq:M}) and $\tilde{M}$ (\ref{eq:tildeM}), we see that
\begin{equation}
\tilde{M} = \gamma M,
\end{equation}
so that $\tilde{M}$, which is defined on the basis of rates depending on the events experienced by the observers, is recognized as the \emph{relativistic mass} in 1+1 dimensions.

\subsection{Action}

The dynamics of a system can be derived by identifying the behavior that minimizes the \emph{action}. Given that the action is such an important quantity in physics, it is interesting that it is not clear precisely what it represents.  In this theory of the free particle, the only observer-independent thing that can be done is simply to count events, and it is to be expected that the action of a particle over an interval is a function of the number of events.  We begin by computing the action of a free particle over the course of $N$ events, which we write as $dN$, in 1+1-dimensions with two observers $\P$ and $\Q$:
\begin{equation}
dS = - E \dt + P \dx,
\end{equation}
where the energy $E$ is given by (\ref{eq:energy;k}), the momentum $P$ is given by (\ref{eq:momentum;k}), and $\dt$ and $\dx$ are given by (\ref{eq:dt:k}) and (\ref{eq:dx:k}), respectively.
With these substitutions, the action is
\begin{multline}
dS = - \frac{1}{4\ddtau}  \left( k + \frac{1}{k} \right) \cdot \frac{1}{2}\left(k \p_0 + \frac{1}{k} \q_0 \right) + \\
\frac{1}{4\ddtau} \left( k - \frac{1}{k} \right) \cdot \frac{1}{2}\left(k \p_0 - \frac{1}{k} \q_0 \right).
\end{multline}
which, given (\ref{eq:dpo}) and (\ref{eq:dqo}), simplifies to
\begin{equation}
dS = -\frac{dN}{2},
\end{equation}
so that the action is an invariant quantity dependent only on the number of events, as one would expect.
The derivation of the action using alternate expressions for energy (\ref{eq:energy;dtau}) and momentum (\ref{eq:momentum;dtau}) greatly simplifies the calculation:
\begin{align}
dS &= - \left( \frac{dN}{2} \frac{\dt}{\dtau^2} \right) \dt + \left( \frac{dN}{2} \frac{\dx}{\dtau^2} \right) \dx \\
&= -\frac{dN}{2} \frac{\left( \dt^2 - \dx^2 \right)}{\dtau^2} \\
&= -\frac{dN}{2}.
\end{align}

The classical action of a relativistic particle evolving from time $t_1$ to $t_2$ along a path $C$ is given by
\begin{align}
S &= - \int\displaylimits_{C}{M \dtau}\\
&= \int_{t_1}^{t_2}{L dt},
\end{align}
in which the \textit{Lagrangian} for the free particle is
\begin{equation}
L = -\frac{M}{\gamma}.
\end{equation}

By writing
\begin{equation}
N = N(p,q),
\end{equation}
the differential $dN$ is
\begin{align}
dN &= \frac{\partial N}{\partial p}dp + \frac{\partial N}{\partial q}dq \\
&= \rp \p + \rq \q \\
&= \sqrt{\rp \; \rq} \left( \sqrt{\frac{\rp}{\rq}} \p + \sqrt{\frac{\rq}{\rp}} \q \right) \\
&= 2M \left( \frac{\frac{1}{k} \p + k \q}{2} \right) \\
&= 2M \dtau.
\end{align}
This illustrates that we can conceive of the mass of a particle in terms of the number of influence events per unit proper time,
\begin{equation}
M = \frac{1}{2}\frac{dN}{\dtau},
\end{equation}
which is the rate at which the particle influences others.

The time derivative of the action $\frac{dS}{dt}$ is
\begin{align}
\frac{dS}{dt} &= \frac{d(-\frac{N}{2})}{dt} = -M \frac{\dtau}{dt}\\
&= -\frac{M}{\gamma},
\end{align}
so that the Lagrangian is
\begin{equation}\label{eq:Lagrangian}
L = \frac{dS}{dt} = -\frac{M}{\gamma},
\end{equation}
as expected.

By (\ref{eq:Lagrangian}) the derivative $\frac{\partial L}{\partial x} = 0$, so that the conjugate momentum,
\begin{align}
P &= \frac{dL}{d\dot{x}}\\
&= M \frac{d}{d\b} \left(-\sqrt{1-\b^2} \right)\\
&= \gamma \b M,
\end{align}
as given in (\ref{eq:P=gbM}), is constant.

\section{The Influenced Particle}
Now that the physics of the free particle is well-understood in this theory, and demonstrated to agree with the kinematics of the free particle in relativistic mechanics, we conclude by briefly considering an \emph{influenced particle}, $\Pi$, which is a particle that is influenced by another.

In this treatment, we will not consider the dynamics of the particle, or particles, that are responsible for influencing the particle $\Pi$.  We begin by imagining a free particle $\Pi$ that influences others at constant rates, which are described as both an energy and momentum.  We then imagine that $\Pi$ is suddenly influenced, relatively infrequently, by another particle at a rate $\rho$.

It is expected that these influence events will interfere with the rates at which the particle influences the observers, and as a result its energy and momentum are expected to change.  The interpretation is that the influenced particle is acted upon by a force (rate of change in momentum).

The \emph{force} $F$ experienced by the particle $\Pi$ is defined as
\begin{equation}
F = \frac{dP}{\dtau}.
\end{equation}
This can be expressed in terms of the Lagrangian by
\begin{align}
\frac{dP}{\dtau} &= \frac{dt}{\dtau} \frac{d}{dt}\left( \frac{\partial L}{\partial \dot{x}} \right)\\
&= \gamma \frac{\partial L}{\partial x} \\
&= - \gamma \frac{\partial U}{\partial x},
\end{align}
for which $U$ represents the \emph{potential energy} and we have used the fact that $L = T - U$.
Since we can also write $P = \gamma \b M = \gamma \dot{x} M$, we have that
\begin{equation}
\frac{dP}{dt} = \gamma^3 M \ddot{x},
\end{equation}
which is the relativistic version of Newton's Second Law in the case for which the acceleration is in the same direction as the velocity of the particle.  It now remains to identify the rate of influence $\rho$ with an energy.

\subsection{Energy Due to Influence}
A free particle, which is not influenced, influences others at a constant outfluence rate, $r = r_\P = r_\Q$, which defines its mass
\begin{equation}
M_0 = \sqrt{r_\P r_\Q} = r.
\end{equation}
In the rest frame, this particle has an energy of
\begin{equation}
E = \frac{r_\P + r_\Q}{2} = r.
\end{equation}
For $N$ outfluence events the proper time experienced by the particle is given by
\begin{equation}
\dtau = N \ddtau,
\end{equation}
so that from (\ref{eq:mass}), a free particle of mass $M_0$ will generate $N$ outfluence events in a proper time of $\dtau$:
\begin{equation}
N = 2 M_0 \dtau.
\end{equation}

In the following sections, we consider two different patterns of influence: a symmetric influence and an anti-symmetric influence.
The symmetric pattern will be shown to be related to a change in energy without an accompanying change in momentum, whereas the anti-symmetric pattern is related to a potential energy distributed in space.  This decomposition is useful since any influence pattern can be expressed as a linear combination of the symmetric and anti-symmetric patterns.

\subsubsection{Symmetric Influence}
We can compare a free particle to an equivalent particle that has been influenced.  In the case of the influenced particle, incoming influence events essentially displace the outfluence events diminishing the rate of outfluence as compared to the free particle.
For a particle influenced symmetrically at a rate of $\rho$ we have that the altered outfluence rates (in comparison to the case of a free particle) are
\begin{align}
{r_\P}' &= r_\P - \frac{\rho}{2}  \label{eq:rP-sym}\\
{r_\Q}' &= r_\Q - \frac{\rho}{2}. \label{eq:rQ-sym}
\end{align}
Since the mass is the geometric mean of the outfluence rates, the effect of influence can be conceived as altering the mass of the particle, $M'$, such that
\begin{align}
{M'}^2 &= {r_\P}'{r_\Q}'\\
&= \left( r_\P - \frac{\rho}{2} \right)  \left( r_\Q - \frac{\rho}{2} \right) \\
&= {M_0}^2 - r_\P \frac{\rho}{2} - r_\Q \frac{\rho}{2} + \frac{\rho^2}{4} \\
&= {M_0}^2 - E \rho + \frac{\rho^2}{4}. \label{eq:m'-squared-1}
\end{align}

The energy of the particle is given by
\begin{equation} \label{eq:E-sym}
E' = E + dE,
\end{equation}
whereas the momentum, which is antisymmetric in the rates (\ref{eq:momentum;r}), remains unchanged due to the symmetry of the influence.
In the rest frame, we have $E = M_0$.  Also, since the momentum does not change, $E' = M'$. By squaring (\ref{eq:E-sym}) we have that
\begin{equation}
{M'}^2 = {M_0}^2 + 2E dE + dE^2. \label{eq:m'-squared-2}
\end{equation}
Equating (\ref{eq:m'-squared-1}) and (\ref{eq:m'-squared-2}), we find that
\begin{equation}
dE = -\frac{\rho}{2}.
\end{equation}
Note that this also follows directly from (\ref{eq:rP-sym}) and (\ref{eq:rQ-sym}) where
\begin{equation}
dE = \frac{{r_\P}' + {r_\Q}'}{2} - \frac{{r_\P} + {r_\Q}}{2}. \label{eq:dE}
\end{equation}
The particle's energy changes, without a change in momentum, since the influence affects the rates in a symmetric fashion.
Because there is no change in momentum, the particle remains at rest in the original rest frame, so that there is no displacement on average and hence no work.
Since a change in internal energy can be written as the difference between the heat transferred to the system and the work done by the system, this change in energy can be considered to be a form of heat.  This surprising result is consistent with the fact that the free particle has an observer-dependent entropy and temperature that are functions of the velocity of the particle \cite{Knuth:MaxEnt2014:motion}.  The entropy of the free particle is a function of the probability that the particle influences $\P$ or $\Q$.  Here, the change in entropy due to the particle being influenced is a function of the probability that the particle is influenced from the $\P$-direction or the $\Q$-direction.  These relationships will be explored in more detail in future work.

\subsubsection{Anti-symmetric Influence}
In this section, we consider an anti-symmetric influence pattern.
Since the energy is symmetric in the rates (\ref{eq:energy;r}), the energy of the system cannot change under an anti-symmetric influence pattern.  This is consistent with energy conservation.  The altered particle outfluence rate is given by
\begin{align}
{r_\P}' &= r_\P + \frac{\rho}{2}  \label{eq:rP-asym} \\
{r_\Q}' &= r_\Q - \frac{\rho}{2}. \label{eq:rQ-asym}
\end{align}
Again, the changes in outfluence rates can be conceived of as an altered mass, $M'$, such that
\begin{align}
{M'}^2 &= {r_\P}'{r_\Q}'\\
&= \left( r_\P + \frac{\rho}{2} \right)  \left( r_\Q - \frac{\rho}{2} \right) \\
&= {M_0}^2 - r_\P \frac{\rho}{2} + r_\Q \frac{\rho}{2} - \frac{\rho^2}{4} \\
&= {M_0}^2 + \rho P - \frac{\rho^2}{4}. \label{eq:m'-squared-3}
\end{align}

Since energy must be conserved in this situation (\ref{eq:energy;r}), the change in potential energy must be equal to the change in mass, which in general (allowing for Lorentz-boosted frames) is
\begin{equation}
\gamma dU = dM,
\end{equation}
so that
\begin{equation}
\gamma dU = \sqrt{{M_0}^2 + \rho P - \frac{\rho^2}{4}} - M_0.
\end{equation}
We can check this as follows.  For cases in which the rates are continuous, $|\rho| \ll M_0$, and when the particle is not hyper-relativistic, $\frac{P}{M_0} \ll \frac{M_0}{\rho}$, and since we are focused on the macroscopic (non-quantum) regime, $P \gg \rho$, we have
\begin{equation}
M' \approx M_0 + \frac{P \rho}{2 M_0}.
\end{equation}
As a result,
\begin{equation}
\gamma dU \approx \frac{P \rho}{2 M_0}.
\end{equation}
Since $dP = -\frac{\rho}{2}$ and $P = \gamma \beta M_0$, we can write
\begin{align}
\gamma dU &\approx - \frac{P dP}{M_0} \\
&\approx -\gamma \beta dP.
\end{align}
Solving for $dP$, we can write
\begin{equation}
dP \approx \frac{-\gamma dU}{\gamma \beta},
\end{equation}
and dividing by $d\tau$ gives
\begin{align}
\frac{dP}{d\tau} &\approx -\frac{\gamma dU}{\gamma \beta d\tau} \\
&\approx -\frac{\gamma dU}{dx},
\end{align}
so that in this case, the influence $\rho$ gives rise to a potential energy.  As expected, the gradient of this potential acts as a relativistic force changing the particle's momentum $P$.

Since any influence affecting the particle can be written as a sum of symmetric and antisymmetric parts,
the effect of a particle being influenced can be expressed as a combination of a change in total energy and a change in potential energy.


\subsection{Potential Energy and Time Dilation}
The change in mass due to potential energy in this context results in time dilation.  Consider a particle $\Pi$ at rest with respect to a pair of coordinated observers $\P$ and $\Q$.  For $N$ outfluence events we have $d\tau = \frac{N}{2M} = dt$.  We will define the potential energy so that for this particle in this position $U = 0$.  We now move the particle to a new position with respect to the observers and bring it to rest so that the new potential energy is $U$, resulting in a change in potential energy $\Delta U = U$.  The particle $\Pi$ has a new mass due to this change in potential energy given by
\begin{align}
M' &= M + dM \\
&= M + \Delta U \\
&= M \left( 1 + \frac{\Delta U}{M} \right).
\end{align}

The mass $M'$ that is measured by the observers is the reciprocal of the length of the projection of an atomic interval from $\Pi$ at rest to either observer so that
\begin{align}
dt &= \frac{N}{2M'} \\
&= \left( \frac{1}{ 1 + \frac{U}{M}} \right) \frac{N}{2M} \,\, ,
\end{align}
whereas the proper time is found by considering the fixed increment $\frac{1}{2M}$ giving
\begin{equation}
d\tau = \frac{N}{2M}.
\end{equation}
As a result, the observer time $dt$ is related to the proper time $d\tau$ by
\begin{equation}
d\tau = \left( 1 + \frac{U}{M} \right) dt.
\end{equation}
In the event that $\frac{U}{M}$ does not depend on any particle properties, all particles experience the same amount of this effect, so that the effect corresponds to what is commonly referred to as time dilation.

\subsection{Influence-Induced Metric}
If we consider a particle that is influenced and has outfluence rates given by (\ref{eq:rP-asym}) and (\ref{eq:rQ-asym}), then the mass of the particle is given by
\begin{align}
{M'}^2 &= {r_\P}'{r_\Q}' \\
&= \frac{N^2}{4 \, dp \; dq}
\end{align}
so that
\begin{align}
{dp \; dq} &= \frac{N^2}{4 {M'}^2} \\
&= \left( 1 + \frac{U}{M} \right)^{-2} \frac{N^2}{4 M^2}.
\end{align}
Since $d\tau = \frac{N}{2M}$ and $dp \; dq = dt^2 - dx^2$ we have that
\begin{equation}
d\tau^2 = \left( 1 + \frac{U}{M} \right)^{2} (dt^2 - dx^2), \label{eq:dtau-U}
\end{equation}
which is a curved metric consistent, in this 1+1-dimensional treatment, with a gravitational potential.  This is consistent with our earlier work \cite{Walsh+Knuth:geodesic} in which geodesic equations were derived corresponding to a metric of the form shown in (\ref{eq:dtau-U}).

\section{Discussion}
Our goal has been to develop a foundational theory in which the familiar concepts of classical mechanics emerge as observer-dependent quantities that describe
the relationships between objects and observers. This work presented here, which we call influence theory, represents an early step toward this goal.  While it would be naive to assume that this will lead to a \textit{final theory}, the progress made thus far is encouraging.  It is our hope that this work will inspire further efforts to develop foundational theories in which the familiar concepts of physics are emergent.

Influence theory considers objects (referred to as particles) that influence one another in a discrete and directed fashion.  This results in a partially ordered set of influence events, which has been shown to be uniquely consistently quantified by the mathematics of relativistic space-time \cite{Knuth+Bahreyni:JMP2014}.  Not only does this model result in a foundation for emergent space-time, but also it results in familiar relativistic kinematics, providing a foundational conception of several quantities central to physics, such as distance, length, duration, mass, energy, and momentum.  These quantities are based either on intervals bounded by events or rates at which events occur, in a way that is reminiscent of de Broglie's electron clock \cite{deBroglie:1924}.
The fact that these quantities represent descriptions of the behavior of particles, rather than properties of particles, is the reason that these quantities are observer-dependent.  This naturally leads to an observer-based physics as a description of the relationship between objects and observers.

Kinematics arises as a changing relationship between object and observer.  The resulting kinematics is based on event-counting, which is demonstrated by the fact that the action of a free particle is proportional to the number of events experienced by the particle.  This formulation of kinematics allows for the development of a Lagrangian formulation, which is a powerful methodology for understanding dynamics, and will be critical for future work.

The dynamics of influenced particles is briefly considered, resulting in a novel perspective of both internal energy and potential energy in terms of influence.  We show that an influenced particle can be treated as a particle that experiences both a change in internal energy (heat) and a change in potential energy with an associated conservative force proportional to its spatial gradient.

In the present context, the potential energy resulting from influence, such that $\frac{U}{M}$ does not depend on any particle properties, results in a time dilation that is in agreement with the form of time dilation due to gravitational forces.  In addition, the behavior of the influenced particle can be described in terms of geodesics and an associated curved space-time.  These results are encouraging when considered in conjunction with the fact that we have demonstrated that inferences about the behavior of a free particle in influence theory give rise to relativistic quantum mechanics \cite{Knuth:Info-Based:2014}.  The fact that we already observe the emergence of physical laws that resemble gravity along with relativistic quantum mechanics suggests that further development of this theory may lead to a framework for quantum gravity.
The results presented here are derived in the context of a pair of coordinated observers, which corresponds to a 1+1-dimensional space-time.

Extending this theory to the case of multiple observers representing a 3+1-dimensional space-time is a current effort.  It is expected that different patterns of influence in higher-dimensional situations will lead to different forces.  While these initial efforts are encouraging, extension to higher dimensions is necessary.  We are hopeful that this first attempt at a foundational theory demonstrates feasibility while providing new insights on how to proceed.

\section*{Acknowledgments}
KHK thanks Newshaw Bahreyni, Ariel Caticha, Seth Chaiken, Keith Earle, Oleg Lunin, Carolyn MacDonald, John Skilling, Matthew Szydagis, and Udo von Toussaint for helpful questions, discussions, and suggestions.  JLW thanks these people and Kevin Vanslette, Philip Goyal, Chris Simmons, Selman Ipek, and Eric Dohner for helpful discussions and suggestions.  KHK especially thanks Hans Eggers, Steve Kroon, and the participants of the 2018 Summer School on Bayesian Inference Foundations and Applications (Chris Engelbrecht Series of Summer Schools associated with Stellenbosch University) in Betty's Bay, South Africa, for the opportunity to present and discuss influence theory, and for their infectious enthusiasm and interest.

\bibliographystyle{unsrt}
\bibliography{C:/Users/Kevinator/kevin/papers/bib/knuth53}

\end{document}